\documentclass[%
 preprint,
 amsmath,amssymb,
 aps,
]{revtex4-1}

\usepackage{graphicx}
\usepackage{dcolumn}
\usepackage{bm}
\usepackage[utf8]{inputenc}
\usepackage{hyperref}
\usepackage{color}

\begin{document}
\title {Force Detection of Electromagnetic Beam Chirality at the Nanoscale}
  
\author{Abid Anjum Sifat$^1$, Filippo Capolino$^1$, Eric O. Potma$^{1,2}$\email{epotma@uci.edu}}
\affiliation{$^1$Department of Electrical Engineering and Computer Science, University of California, Irvine}
\affiliation{$^2$Department of Chemistry, University of California, Irvine}





\begin{abstract}
Many nanophotonic applications require precise control and characterization of electromagnetic field properties at the nanoscale. The chiral properties of the field are among its key characteristics, yet measurement of optical chirality at dimensions beyond the diffraction limit has proven difficult. Here we theoretically show that the chiral properties of light can be characterized down to the nanometer scale by means of force detection. To measure the chiral properties of a beam of given handedness at the nanoscale, we determine the photo-induced force exerted on a sharp tip, which is illuminated first by the beam of interest and second by an auxiliary beam of opposite handedness, in a sequential manner. We show that the difference between the force measurements is directly proportional to the chiral properties of the beam of interest. In particular, the gradient force difference $\Delta\langle$\textit{$F_{grad, z}$}$\rangle$ is found to have exclusive correspondence to the time-averaged helicity density of the incident light, whereas the differential scattering force provides information about the spin angular momentum density of light. We further characterize and quantify the helicity-dependent $\Delta\langle$\textit{$F_{grad, z}$}$\rangle$ using a Mie scattering formalism complemented with full wave simulations, underlining that the magnitude of the difference force is within an experimentally detectable range.
\end{abstract}

\maketitle
\section{Introduction}

Light can exist in a chiral state, and this property makes it possible to study chiral objects, such as molecules and nanostructures, in an optical manner.\cite{tang2011enhanced,zhao2016enantioselective,gorkunov2017enhanced,govorov2010theory,govorov2011plasmon,amabilino2009chirality} Optical detection of chirality is important in a variety of scientific fields, with key applications in molecular biology and pharmacology.\cite{behr2008lock,denmark2006topics,tverdislov2017periodic,kasprzyk2010pharmacologically} In common applications, the specimen is illuminated sequentially with left-handed and right-handed circularly polarized light (LCP and RCP), and the chiral information is inferred from the differential response of the material. The chiral state of circularly polarized light is well understood, which facilitates the analysis of such measurements. However, performing similar measurements at the nanoscale can be challenging~\cite{ho2017enhancing,hanifeh2020helicity,hanifeh2020helicitymax}, because the chiral state of light in the near zone, close to a nano-object, is generally unknown and cannot be directly assessed experimentally. Without experimental tools for extracting information about the chirality of the electromagnetic field, optical examination of chirality at the nanoscale is severely complicated. 

The chiral state of light is related to the curled character of the electric and magnetic fields. The 
quantity commonly used for quantifying this `degree of curliness' is the time-averaged helicity density \cite{bliokh2013dual,cameron2012optical,hanifeh2020helicitymax}, which is defined as {$h=\frac{1}{2\omega c}$}\rm{Im}($\mathbf{E}\cdot\mathbf{H^{\ast}}$), where $\mathbf{E}$ and $\mathbf{H}$ are the phasor electric and magnetic field at angular frequency $\omega$, and $c$ is the speed of light ($^\ast$ denotes complex conjugation) \cite{trueba1996electromagnetic,hanifeh2020optimally}. The helicity density is a time-even, pseudo-scalar conserved quantity; when using CP light, it corresponds to the difference between the numbers of RCP and LCP photons \cite{bliokh2013dual,cameron2012optical,bliokh2011characterizing}. The flux of the helicity density, i.e. the chiral momentum density, is related to the spin angular momentum of the beam  \cite{bliokh2013dual,bliokh2011characterizing}. The latter quantity is a time-odd, pseudo-vector quantity \cite{cameron2012optical} and its density is defined as ${\boldsymbol\sigma}=-\frac{\varepsilon_0}{4\omega i}$($\mathbf{E}\times\mathbf{E^{\ast}})-\frac{\mu_0}{4\omega i}$($\mathbf{H}\times\mathbf{H^{\ast}}$), where $\varepsilon_0$ and $\mu_0$ are the absolute permittivity and permeability of free space  \cite{cameron2012optical,barnett2010rotation}, respectively. Although both quantities describe angular momentum associated with the polarization state of light\cite{bliokh2013dual}, only the time-averaged helicity density is classified as a conserved property of the electromagnetic field \cite{tang2010optical,tang2011enhanced}, and, therefore, considered to be the proper descriptor of optical chirality.

The question arises whether a quantity like the helicity density can be properly measured at the nanoscale. Unlike quantities such as energy, mechanical force and torque carried or performed by the electromagnetic field, properties like linear and angular momentum of light are more difficult to measure and are commonly deduced from other measurable quantities \cite{emile2018energy}. Helicity density of light falls into the latter category, and measuring it requires a connection to a quantity that can be experimentally assessed. One possibility is to detect helicity density through a photo-induced force. For instance, the technique of photo-induced force microscopy (PiFM) has been used to map electric and magnetic field distributions through the optical force exerted on a metallic tip \cite{zeng2018sharply} or a specially designed magnetic nanoprobe \cite{zeng2021photoinduced}. The photo-induced force has also been used for sorting and trapping of chiral nanoparticles\cite{hou2021separating,kamenetskii2021chirality,hayat2015lateral,tkachenko2014optofluidic,li2019optical}. In addition, PiFM has been employed to determine enantioselective chirality of nanosamples \cite{kamandi2017enantiospecific}, for measuring the geometric chirality of broken symmetry structures with differential force measurements under RCP/LCP illumination\cite{rajaei2019giant} and for measuring enantioselective optical forces as well\cite{zhao2017nanoscopic} .

In this work, we theoretically investigate the connection between the photo-induced forces felt by a chiral tip under differential RCP/LCP illumination and the chiral state of light at the nanoscale. Using image dipole theory, we have found a direct relation between the measured differential force and the chiral properties of the incident electromagnetic field in terms of the time-averaged helicity density and spin-angular momentum density.  The $\exp(-i \omega t)$ time notation is implicitly assumed throughout the paper. We further model the chiral tip as an isotropic chiral sphere to examine several design considerations and validate our dipole model with full-wave, finite element method (FEM) simulations.


\section{Theoretical Analysis}\label{sec:point_dipole}
\begin{figure}[h]
\centering
\includegraphics[height=5cm]{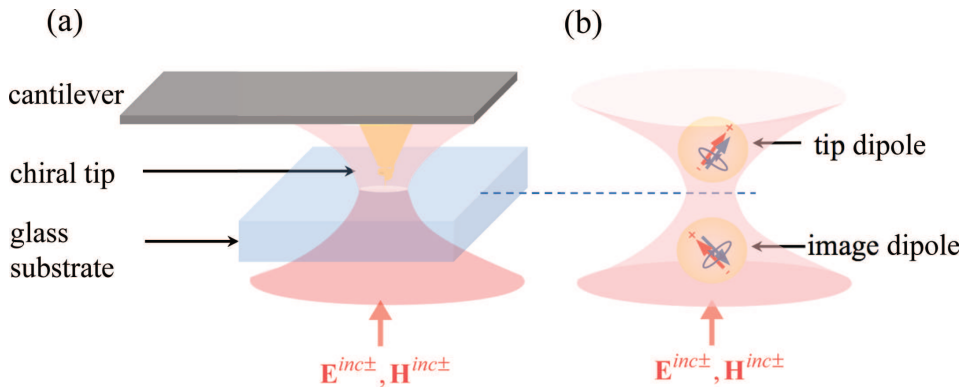}
\caption{(a) Schematic of the PiFM system with a chiral tip with illumination from the bottom. (b) Photo-induced chiral tip dipole and image dipole according to image dipole theory. }
\label{fig:fig1}
\end{figure}

We are interested in detecting the mechanical force that acts on the chiral tip due to the presence of  electromagnetic field. We can model the tip as particle and write the time-averaged force exerted on this particle as \cite{novotny2012principles, yang2016resonant}
\begin{equation}\label{eq:Force_general}
    \langle\mathbf{F}\rangle=\frac{1}{2}
\mathrm{Re}\Big\{\int\limits_{\mathit{S}}
\left[\varepsilon(\mathbf{E}\cdot\hat{\mathbf{n}})\mathbf{E}^{\ast}
+\mu^{-1}(\mathbf{B}\cdot\hat{\mathbf{n}})\mathbf{B}^{\ast}
-\frac{1}{2}\left(\varepsilon|\mathbf{E}|^2+\mu^{-1}|\mathbf{B}|^2\right)\hat{\mathbf{n}}\right]
\mathit{dS}\Big\}
\end{equation}
where the integration is over the arbitrary surface $S$ that encloses the tip-particle of volume $V$, $\hat{\mathbf{n}}$ is the unit vector normal to this surface, and $\mathbf{E}$ and $\mathbf{B}$ are the phasors of the total electric field and magnetic flux density, which include both the incident light and the scattered light contributions. It is assumed that the particle is embedded in a non-dissipative medium with permittivity {$\varepsilon$} and permeability $\mu$. To simplify this expression, we next assume that the tip can be described as a dipolar particle with an electric and magnetic dipole moment in free space, written as $\mathbf{p}_{tip}~(\rm{Cm})$ and $\mathbf{m}_{tip}~(\rm{Am}^2)$, respectively. The phasors of the dipole moments are defined as $\mathbf{p}_{tip}=\int \mathbf{r}\rho(\mathbf{r}) dV$ and $\mathbf{m}_{tip}=\frac{1}{2}\int [\mathbf{r}\times\mathbf{J}(\mathbf{r})] dV$, where $\rho(\mathbf{r})$ and $\mathbf{J}(\mathbf{r})$ are the induced charge and current densities in the particle. Assuming the tip-particle's response is dominated by its electric and magnetic dipole moments, the expression for the time-averaged electromagnetic force reduces to  
\cite{nieto2010optical,yaghjian1999electromagnetic} 

\begin{equation}\label{eq:dipolar_force}
\begin{split}
\langle\mathbf{F}\rangle=\frac{1}{2}\mathrm{Re}\Big\{
\big(\nabla\mathbf{E}^{loc}(\mathbf{r}_{tip})\big)^\ast\cdot\mathbf{p}_{tip}
+\big(\nabla\mathbf{H}^{loc}(\mathbf{r}_{tip})\big)^\ast\cdot\mu_0\mathbf{m}_{tip} \\
-\frac{c\mu_0 k^4}{6\pi}(\mathbf{p}_{tip}\times\mathbf{m}^{\ast}_{tip})\Big\}
\end{split}
\end{equation}
 where $\mathbf{E}^{loc}(\mathbf{r}_{tip})$ and $\mathbf{H}^{loc}(\mathbf{r}_{tip})$ are the local electric and magnetic field at the tip dipole position. Here, $\nabla\mathbf{E}$ and $\nabla\mathbf{H}$ are the gradient of electric and magnetic field vectors. This is of dyadic form and defined as $\nabla\mathbf{E}=\sum_{i} \sum_{j} \frac{\partial E_j}{\partial x_i}\hat{x_i}\hat{x_j}$ where $i$ and $j$ stands for $x$, $y$ and $z$ components of cartesian coordinates.\cite{tai1994dyadic} Note that sometimes in the literature the $\mu_0$ factor in the second and third term of equation (\ref{eq:dipolar_force}) is included in the definition of the magnetic dipole moment ($\mathbf{m}_{tip}$).\cite{Wang2014,kamandi2017enantiospecific, kamandi2018unscrambling} The notation used in this paper is the same as that in Refs. \cite{zeng2018exclusive,hanifeh2020helicity,hanifeh2020helicitymax,hanifeh2020optimally,zeng2021photoinduced}.  The time-averaged force has three distinct contributions: the first term on the right hand side of equation (\ref{eq:dipolar_force}) is recognized as the electric dipolar force $\langle\mathbf{F}_{e}\rangle$, the second term is known as the magnetic dipolar force $\langle\mathbf{F}_{m}\rangle$, and the last term is called the interaction force $\langle\mathbf{F}_{int}\rangle$. 

To calculate the force, we require expressions for the induced electric and magnetic dipole moments. Assuming that the tip is an isotropic and reciprocal chiral object, the dipole moments can be related to the local fields through polarizabilities as \cite{A} 

\begin{align}
  \mathbf{p}_{tip}&=\alpha_{ee}\mathbf{E}^{loc}(\mathbf{r}_{tip})+\alpha_{em}\mathbf{H}^{loc}(\mathbf{r}_{tip})\label{eq:electric_dipole_tip}\\
  \mathbf{m}_{tip}&=-\mu^{-1}_0\alpha_{em}\mathbf{E}^{loc}(\mathbf{r}_{tip})+\alpha_{mm}\mathbf{H}^{loc}(\mathbf{r}_{tip})\label{eq:magnetic_dipole_tip}
\end{align}
where $\alpha_{ee}$, $\alpha_{mm}$ and $\alpha_{em}$ are the electric, magnetic and electro-magnetic polarizabilities, respectively. We have considered that $\alpha_{me}=- \mu_0^{-1} \alpha_{em}$  because of reciprocity. The local fields at the tip dipole position can be obtained from image dipole theory \cite{novotny2012principles, jackson2009classical,rajapaksa2010image}. We model the effect of the substrate by substituting the substrate response by the image of the photo-induced (chiral) dipole in the tip, written as $\mathbf{p}_{img}$ and $\mathbf{m}_{img}$ and shown in Figure \ref{fig:fig1}(b). The local electric and magnetic fields at the tip position can then be expressed as 
\begin{align}
    \mathbf{E}^{loc}(\mathbf{r}_{tip})&=\mathbf{E}^{inc}(\mathbf{r}_{tip})+\mathbf{E}^{sca}_{img\rightarrow tip}(\mathbf{r}_{tip})\label{eq:E_loc}\\
    \mathbf{H}^{loc}(\mathbf{r}_{tip})&=\mathbf{H}^{inc}(\mathbf{r}_{tip})+\mathbf{H}^{sca}_{img\rightarrow tip}(\mathbf{r}_{tip})\label{eq:H_loc}
\end{align}
where $\mathbf{E}^{inc}(\mathbf{r}_{tip})$, $\mathbf{H}^{inc}(\mathbf{r}_{tip})$ are the incident electric and magnetic field at the tip dipole location, and $\mathbf{E}^{sca}_{img\rightarrow tip}(\mathbf{r}_{tip})$, $\mathbf{H}^{sca}_{img\rightarrow tip}(\mathbf{r}_{tip})$ are the scattered electric and magnetic field by the image dipoles at the tip dipole location.

The scattered fields, which include nearfields, are defined through the image dipole moments and their Green’s functions as  \cite{novotny2012principles,jackson2009classical} 
\begin{align}
    \mathbf{E}^{sca}_{img\rightarrow tip}(\mathbf{r}_{tip})&=\mathbf{\underline{G}}^{ee}(\mathbf{r}_{tip}-\mathbf{r}_{img})\cdot\mathbf{p}_{img}+\mathbf{\underline{G}}^{em}(\mathbf{r}_{tip}-\mathbf{r}_{img})\cdot\mathbf{m}_{img}\label{eq:Esca}\\
    \mathbf{H}^{sca}_{img\rightarrow tip}(\mathbf{r}_{tip})&=\mathbf{\underline{G}}^{me}(\mathbf{r}_{tip}-\mathbf{r}_{img})\cdot\mathbf{p}_{img}+\mathbf{\underline{G}}^{mm}(\mathbf{r}_{tip}-\mathbf{r}_{img})\cdot\mathbf{m}_{img}\label{eq:Hsca}
\end{align}
where $\mathbf{\underline{G}}(\mathbf{r})=G_x\mathbf{\hat{x}\hat{x}}+G_y\mathbf{\hat{y}\hat{y}}+G_z\mathbf{\hat{z}\hat{z}}$ is the dyadic Green’s function in Cartesian coordinates. The  Green's functions that appear in equations (\ref{eq:Esca}) and (\ref{eq:Hsca}) are approximated in the nearfield as 
\cite{campione2013effective} 
\begin{align}
    \mathbf{\underline{G}}^{ee}(\mathbf{r})&=\frac{1}{4\pi\varepsilon_0|\mathbf{r}|^{3}}(3\mathbf{\hat{r}}\mathbf{\hat{r}}-\mathbf{\underline{I}})\\
    \mathbf{\underline{G}}^{mm}(\mathbf{r})&=\frac{1}{4\pi|\mathbf{r}|^{3}}(3\mathbf{\hat{r}}\mathbf{\hat{r}}-\mathbf{\underline{I}})\\
    \mathbf{\underline{G}}^{em}(\mathbf{r})&=-\frac{1}{i\omega\varepsilon_0}\nabla\times\mathbf{\underline{G}}^{mm}(\mathbf{r})\\
    \mathbf{\underline{G}}^{me}(\mathbf{r})&=\frac{1}{i\omega\mu_0}\nabla\times\mathbf{\underline{G}}^{ee}(\mathbf{r})
\end{align}
where $\mathbf{\underline{I}}$ is unit dyad (tensor) and $\mathbf{\hat{r}}$ is the unit vector from the source point to the point of observation.

In the same fashion, we can also define the image dipole moments and the local fields at the image location. Our goal is to obtain self-consistent approximate expressions for the local fields at the tip location and  for the dipole moments in terms of the incident light components. To achieve this, consistent with the nearfield approximation of the Green's function, we first ignore any phase retardation effects in the fields between the tip and image dipole position. Second, we will only consider terms in the polarizability up to second order (full derivation is provided in the Supplementary Information). With these assumptions, we may use equations (\ref{eq:electric_dipole_tip})-(\ref{eq:H_loc}) to determine the longitudinal ($z$) component of the time-averaged force in the dipole approximation as written as
\begin{equation}
\begin{split}
\langle{F_{z,grad}}\rangle=\langle{F_{z,e}}\rangle+\langle{F_{z,m}}\rangle\\
=\frac{3}{2\pi|\mathit{z}|^4}~&\Bigg[\frac{1}{2}\bigg(-\frac{|\alpha_{ee}|^2}{\varepsilon_0}+\mu_0|\mu^{-1}_0\alpha_{em}|^2\bigg)(\tfrac{1}{2}|\mathbf{E}^{inc}_\parallel|^2+|\mathbf{E}^{inc}_z|^2)\\
&+\frac{1}{2}\bigg(-\frac{|\alpha_{em}|^2}{\varepsilon_0}+\mu_0|\alpha_{mm}|^2\bigg)(\tfrac{1}{2}|\mathbf{H}^{inc}_\parallel|^2+|\mathbf{H}^{inc}_z|^2)\\
&-\mathrm{Re}\bigg\{\bigg(\frac{\alpha_{ee}\alpha^{\ast}_{em}}{\varepsilon_0}+\alpha^{\ast}_{mm}\alpha_{em}\bigg)
(\tfrac{1}{2}\mathbf{E}^{inc}_{\parallel}\cdot\mathbf{H}^{inc\ast}_{\parallel}+\mathbf{E}^{inc}_{z}\cdot\mathbf{H}^{inc\ast}_{z})\bigg\}\Bigg]\\
+\frac{6\omega}{\pi k^{2}|\mathit{z}|^5}&\Bigg[2\omega\mathrm{Re}\bigg\{\frac{{\alpha_{ee}\alpha^{\ast}_{em}}}{\varepsilon_0}\bigg\}[\boldsymbol{\sigma}^{inc}_{E}]_{z}-2\omega\mathrm{Re}\big\{{\alpha_{em}\alpha^{\ast}_{mm}}\big\}[\boldsymbol{\sigma}^{inc}_{H}]_{z}\\
&-2\mathrm{Im}\Big\{(|\alpha_{em}|^{2}-\mu_{0}\alpha_{ee}\alpha_{mm}^{\ast})[\mathbf{S}^{inc}]_{z}\Big\}\Bigg]
\end{split}
\label{eq:force_em}
\end{equation}

\begin{equation}
\begin{split}
\langle{F_{z,int}}\rangle=~&-\frac{ck^{4}}{12\pi}\Bigg[\mathrm{Re}\Big\{-\alpha^{\ast}_{em}\alpha_{ee}\Big\}\mathrm{Re}\Big\{[\mathbf{E}^{inc}_{\parallel}\times\mathbf{E}^{inc\ast}_{\parallel}]_{z}\Big\}\\
&+\mathrm{Re}\Big\{\mu_0\alpha^{\ast}_{mm}\alpha_{em}\Big\}\mathrm{Re}\Big\{[\mathbf{H}^{inc}_{\parallel}\times\mathbf{H}^{inc\ast}_{\parallel}]_{z}\Big\}\\
&+4\omega\mathrm{Im}\Big\{-\frac{\alpha^{\ast}_{em}\alpha_{ee}}{\varepsilon_0}\Big\}[\boldsymbol{\sigma}^{inc}_{E}]_{z}+4\omega\mathrm{Im}\Big\{-\alpha^{\ast}_{mm}\alpha_{em}\Big\}[\boldsymbol{\sigma}^{inc}_{H}]_{z}\\
&+2\mathrm{Re}\Big\{\alpha^{\ast}_{em}\alpha_{em}[\mathbf{S}^{inc\ast}]_{z}\Big\}+2\mathrm{Re}\Big\{\mu_0\alpha^{\ast}_{mm}\alpha_{ee}[\mathbf{S}^{inc}]_{z}\Big\}\Bigg]
\end{split}
\label{eq:force_int}
\end{equation}
where $\mathbf{E}^{inc}_{\parallel}=\mathit{E}^{inc}_x\mathbf{\hat{x}}+\mathit{E}^{inc}_y\mathbf{\hat{y}}$,~$\mathbf{H}^{inc}_{\parallel}=\mathit{H}^{inc}_x\mathbf{\hat{x}}+\mathit{H}^{inc}_y\mathbf{\hat{y}}$ and $\mathbf{E}^{inc}_{z}=\mathit{E}^{inc}_z\mathbf{\hat{z}}$, $\mathbf{H}^{inc}_{z}=\mathit{H}^{inc}_z\mathbf{\hat{z}}$ are the transverse and longitudinal components of the incident fields at the tip dipole location, the notation $(\mathbf{r}_{tip})$ has been avoided here for brevity; $\mathbf{S}^{inc}=\frac{1}{2}\mathbf{E}^{inc}\times(\mathbf{H}^{inc})^\ast$ is the Poynting vector of the incident field; $\boldsymbol{\sigma}^{inc}_{E}$ and $\boldsymbol{\sigma}^{inc}_{H}$ are the electric and magnetic parts of the time-averaged total spin angular momentum  density of the incident light; and $z$ is the vertical distance between the tip dipole and its image. See the Supporting Information for a detailed derivation of equations (\ref{eq:force_em}) and (\ref{eq:force_int}).

Equation (\ref{eq:force_em}) describes the force due to the combined electric and magnetic dipolar response of the tip. This force shows a distance dependence on the tip-sample distance which scales as $\mathit{z}^{-4}$, similar to the distance dependence of the gradient force that is typically measured in PiFM \cite{novotny2012principles,jahng2014gradient}. The first two lines of equation (\ref{eq:force_em}) are recognized as the purely electric and purely magnetic dipolar force contributions to the gradient force, whereas the third line describes the force that arises from a nonzero scalar product of the electric and magnetic fields, i.e. the helicity density.  In addition, this force also exhibits a $z^{-5}$ distance dependence, expressed in the last two lines of equation (\ref{eq:force_em}). This part carries information about the longitudinal component of the spin angular momentum density and the Poynting vector of the incident light.

Equation (\ref{eq:force_int}) accounts for the interaction force, which lacks a direct dependence on the tip-sample distance. In addition to the purely electric and magnetic contributions to the interaction force, described by the first two lines in equation (\ref{eq:force_int}), the latter two lines add force contributions that scale with the longitudinal component of the spin angular momentum density and with the Poynting vector. Because these latter terms depend on the momentum of the incoming field, the interaction force shows similarity with the scattering force \cite{novotny2012principles,jahng2014gradient} in PiFM. 

We are interested in using the force as a way to measure the helicity density $h^{inc}$ of an incident beam. To measure this quantity, we will use a second incident beam that is similar to the original beam but is of opposite handedness. Hence, if we define the helicity density of the original beam as $h^{inc}=h^{inc+}$, then the helicity density of the auxiliary beam is given as $h^{inc-}$. For this purpose, we assume two states for the incident light, indicated by $\mathbf{E}^{inc+}$, $\mathbf{H}^{inc+}$ and $\mathbf{E}^{inc-}$, $\mathbf{H}^{inc-}$, describing the input fields of different handedness. The two illumination states have the same energy densities, $|\mathbf{E}^{inc+}|^2=|\mathbf{E}^{inc-}|^2$ and $|\mathbf{H}^{inc+}|^2=|\mathbf{H}^{inc-}|^2$, but exhibit opposite helicity densities, $\mathrm{Im}(\mathbf{E}^{inc+}\cdot\mathbf{H}^{inc+\ast})=-\mathrm{Im}(\mathbf{E}^{inc-}\cdot\mathbf{H}^{inc-\ast})$, i.e. $h^{inc+}=-h^{inc-}$. Note that the longitudinal spin angular momentum density components are related as $[\boldsymbol{\sigma}^{inc+}_{E}]_z=-[\boldsymbol{\sigma}^{inc-}_{E}]_z$ and $[\boldsymbol{\sigma}^{inc+}_{H}]_z=-[\boldsymbol{\sigma}^{inc-}_{H}]_z$. We also find that $\mathbf{S}^{inc+}=\mathbf{S}^{inc-}$, and that the following relations hold

\begin{equation}
    \begin{split}
        \mathrm{Re}\Big\{[\mathbf{E}^{inc+}_{\parallel}\times\mathbf{E}^{inc+\ast}_{\parallel}]_{z}\Big\}&=\mathrm{Re}\Big\{[\mathbf{E}^{inc-}_{\parallel}\times\mathbf{E}^{inc-\ast}_{\parallel}]_{z}\Big\}\\
        \mathrm{Re}\Big\{[\mathbf{H}^{inc+}_{\parallel}\times\mathbf{H}^{inc+\ast}_{\parallel}]_{z}\Big\}&=\mathrm{Re}\Big\{[\mathbf{H}^{inc-}_{\parallel}\times\mathbf{H}^{inc-\ast}_{\parallel}]_{z}\Big\}
    \end{split}
\end{equation}

We next determine the differential force, obtained by measuring the force under illumination with incident light of $(+)$ and $(-)$ handedness, and taking the difference. Under these conditions, the differential gradient force $\Delta\langle{F_{z,grad}}\rangle=\langle{F_{z,grad}}\rangle^+-\langle{F_{z,grad}}\rangle^-$ and the differential interaction force $\Delta\langle{F_{z,int}}\rangle=\langle{F_{z,int}}\rangle^+-\langle{F_{z,int}}\rangle^-$ can be obtained from equations (\ref{eq:force_em}) and (\ref{eq:force_int}) as
\begin{equation}
\begin{split}
\Delta\langle{F_{z,grad}}\rangle=~&\frac{6\omega c}{\pi|\mathit{z}|^4}
\mathrm{Im}\Big\{\frac{\alpha^{\ast}_{em}}{\varepsilon_0}(\alpha_{ee}-\alpha_{mm}\varepsilon_0)\Big\} \Big(\tfrac{1}{2}\mathit{h}^{inc}_{\parallel}+\mathit{h}^{inc}_{z}\Big)\\
+&\frac{24\omega^2}{\pi k^{2}|\mathit{z}|^5}\Bigg[\mathrm{Re}\left\{\frac{\alpha_{ee}\alpha^{\ast}_{em}}{\varepsilon_o}\right\}[\boldsymbol{\sigma}^{inc}_{E}]_{z}-\mathrm{Re}\left\{{\alpha_{mm}\alpha^{\ast}_{em}}\right\}[\boldsymbol{\sigma}^{inc}_{H}]_{z}\Bigg]\end{split} \label{eq:force_grad}
\end{equation}
\begin{equation}
\begin{split}
\Delta\langle{F_{z,int}}\rangle&=\frac{2\omega ck^4}{3\pi}\Big[
\mathrm{Im}\Big\{\frac{\alpha^{\ast}_{em}\alpha_{ee}}{\varepsilon_o}\Big\}[\boldsymbol{\sigma}^{inc}_{E}]_{z}+\mathrm{Im}\Big\{\alpha_{mm}\alpha^{\ast}_{em}\Big\}[\boldsymbol{\sigma}^{inc}_{H}]_{z}\Big]
\end{split} 
\label{eq:force_scat}
\end{equation}
where $\mathit{h}^{inc}_{\parallel}=\frac{1}{2\omega c}\mathrm{Im}\left\{\mathbf{E}^{inc}_{\parallel}\cdot\mathbf{H}^{inc\ast}_{\parallel}\right\}$ and $\mathit{h}^{inc}_{z}=\frac{1}{2\omega c}\mathrm{Im}\left\{\mathbf{E}^{inc}_{z}\cdot\mathbf{H}^{inc\ast}_{z}\right\}$ are the tangential and longitudinal helicity density, respectively. We find that the differential force is proportional to the difference in helicity density between the incident light and the auxiliary beam, i.e. $h^{inc+}-h^{inc-}=2h^{inc}$.  Since we had defined $h^{inc}=h^{inc+}$, we observe that the differential gradient force measured in this procedure is directly proportional to the desired $h^{inc}$ of the incident chiral light. From equation (\ref{eq:force_grad}) we see that the differential gradient force depends on the helicity density of the incident light, whereas (\ref{eq:force_scat}) predicts that the differential interaction force carries only information on the spin angular momentum of the incident light. When using optimally chiral light\cite{hanifeh2020optimally} satisfying the relation $\mathbf{E}=\pm i \eta_0 \mathbf{H}$ with $\eta_0=\sqrt{\mu_0/{\varepsilon_0}}$, the two spin angular momentum densities are such that $\boldsymbol{\sigma}^{inc}_{H} =\boldsymbol{\sigma}^{inc}_{E}$. Note that circularly polarized light is a particular case of optimally chiral light. 

When analyzing the first and the second terms in equation (\ref{eq:force_grad}), we note that the pre-factor containing $z^{-5}$ grows larger than the corresponding $z^{-4}$ pre-factor for sub-wavelength distances. The magnitude of the two contributions to the differential gradient force are, however, also determined by the material polarizabilities. In this context, we must observe that for nanoparticles made of low loss materials like silicon, the values for the polarizabilities $\alpha_{ee}$ and $\alpha_{mm}$ are mainly real when the particle is much smaller than the optical wavelength (i.e., at the quasistatic limit). Assuming the same particle has chirality, we also observe that $\alpha_{em}$ is mainly imaginary (see for example the expression (9) in Ref. \cite{kamandi2017enantiospecific} or (7) in Ref.\cite{kamandi2018unscrambling}). Therefore, when the chiral tip is made of a low-loss material and much smaller than the optical wavelength, the term $\mathrm{Im}\left\{\alpha_{ee}\alpha^{\ast}_{em}\right\}$ is larger than the term $\mathrm{Re}\left\{\alpha_{ee}\alpha^{\ast}_{em}\right\}$. Similarly, we find that $\mathrm{Im}\left\{\alpha_{mm}\alpha^{\ast}_{em}\right\}\gg\mathrm{Re}\left\{\alpha_{mm}\alpha^{\ast}_{em}\right\}$. 
Taken together, even though the considered tip size is not deeply sub-wavelength, we find that for practical sizes and material properties of the tip, as shown in Section \ref{sec:FWS}  (see also Supporting Information section 6), the first term in equation (\ref{eq:force_grad}) is dominant over the second term. For this reason,  we may approximate the differential gradient force as 
\begin{equation}
\begin{split}
\Delta\langle{F_{z,grad}}\rangle\sim~&\frac{6\omega c}{\pi|\mathit{z}|^4}
\mathrm{Im}\Big\{\frac{\alpha^{\ast}_{em}}{\varepsilon_0}(\alpha_{ee}-\alpha_{mm}\varepsilon_0)\Big\} \Big(\tfrac{1}{2}\mathit{h}^{inc}_{\parallel}+\mathit{h}^{inc}_{z}\Big)\end{split} \label{eq:force_grad_approx}
\end{equation}
and the validity of such approximation will be confirmed by the numerical results. It is clear that the strength and sign of the differential gradient force depends on $\alpha^{\ast}_{em}$ and thus relies on the chiral properties of the tip. For small tip-sample distances, the differential gradient force is expected to constitute the dominant contribution to the measured force, thus offering a means of experimentally extracting information about the helicity density. Unless otherwise stated, we will use equation (\ref{eq:force_grad_approx}) to determine the differential force.


\section{Mie Scattering Formalism}

We next study the characteristics of the differential gradient force.  We see from equation (\ref{eq:force_grad_approx}) that $\Delta\langle{F_{z,grad}}\rangle$ depends on the polarizability of the tip material, including the electro-magnetic polarizability. To model the tip polarizability, we assume that the tip can be described as a spherical chiral nanoparticle (NP). We also assume that the following constitutive relations hold:  $\mathbf{D}=\varepsilon_0\varepsilon_r\mathbf{E}+i\sqrt{\varepsilon_0\mu_0}\kappa\mathbf{H}$ and $\mathbf{B}=\mu_0\mu_r\mathbf{H}-i\sqrt{\varepsilon_0\mu_0}\kappa\mathbf{E}$, where $\varepsilon_r$ and $\mu_r$ are the relative permittivity and relative permeability, respectively.\cite{A,hanifeh2020optimally} The chirality parameter $\kappa$ is an empirical quantity that provides the chiral strength of the material under consideration. We can next relate the electric, magnetic and electro-magnetic polarizabilities through the material parameters by using Mie scattering theory as:\cite{kamandi2017enantiospecific,bohren2012univ,hanifeh2020optimally}  $\alpha_{ee}=-{6\pi i \varepsilon_0 b_1}/{k^3_0}$, $\alpha_{mm}=-{6\pi i a_1}/{k^3_0}$ and $\alpha_{em}={6\pi i c_1}/{(c k^3_0)}$ where $c$ is the free space speed of light, $k_0$ is the wavenumber in free space and $b_1$, $a_1$, $c_1$ are the Mie coefficients. In our calculations, the material parameters of the isotropic chiral NP such as $\varepsilon_r$ and $\mu_r$ are those of crystalline silicon\cite{aspnes1983dielectric} with $\mu_r=1$.

\begin{figure}[h]
\centering
\includegraphics[width=7cm]{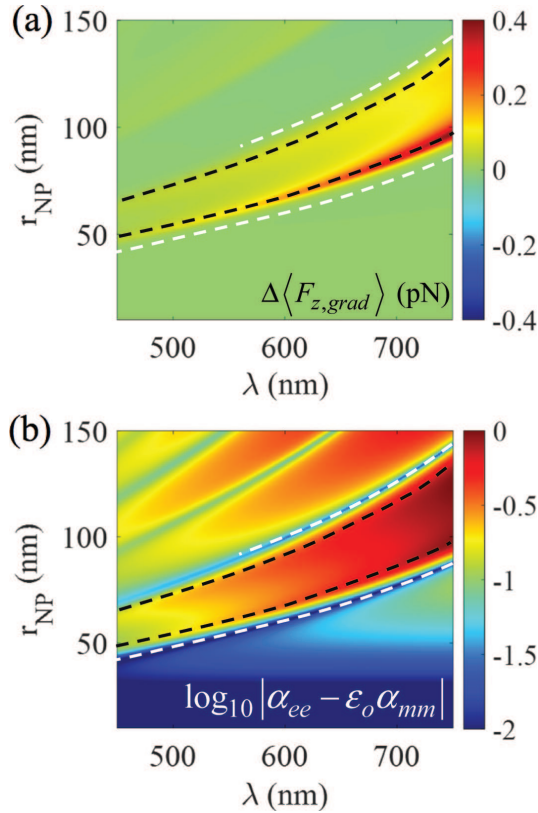}
\caption{(a) Differential gradient force spectrum in pN. (b) Normalized magnitude spectrum of $\alpha_{ee}-\varepsilon_0\alpha_{mm}$ for different radii of the chiral NP. The material properties of the chiral NP, $\varepsilon_r$ and $\mu_r$ are considered the same as silicon with $\mu_r=1$ and $\kappa=0.1$. In all calculations, the tip-image dipole center-center distance is $|z|=2r_{NP}+10~\rm{nm}$. Black dashed lines indicate resonance condition of $c_1$ and white dashed lines indicate the local minima in the plot, which approach the  Kerker condition. }
\label{fig:fig2}
\end{figure}

We next place the sphere (i.e., the tip) just above a flat and transparent dielectric substrate  such that the distance between the sphere's surface and the substrate surface is 5~nm. The system is subsequently illuminated from the bottom with a plane wave of field strength $1.5\times10^6 ~\rm{Vm}^{-1}$ that is either in the LCP ($\mathbf{E}^{inc+}=\mathit{E}^{inc}\mathbf{\hat{x}}+i\mathit{E}^{inc}\mathbf{\hat{y}}$) or RCP ($\mathbf{E}^{inc-}=\mathit{E}^{inc}\mathbf{\hat{x}}-i\mathit{E}^{inc}\mathbf{\hat{y}}$) state, and which propagates through the transparent dielectric material toward the sphere. The CP wave induces  an electric and magnetic dipole in the chiral sphere, and the presence of the substrate can subsequently be modeled by including an image of the induced electric and magnetic  dipoles\cite{zeng2018sharply,zeng2021photoinduced}, analogous to the procedure used for the point dipole model in section \ref{sec:point_dipole}. For simplicity, it is assumed that the image dipole strength is identical to the induced dipole in the sphere. In the current configuration, the distance between the location of the dipole and its image is  $2(r_{NP}+5~\rm{nm})$, where $r_{NP}$ is the radius of the sphere. We calculate the differential gradient force spectrum with the aid of equation (\ref{eq:force_grad_approx}) and plot it as a function of sphere radius and excitation wavelength in Fig. \ref{fig:fig2}(a). In all calculations, the value of the bulk chirality $\kappa$ of the sphere is  considered real and set to 0.1.\cite{ali2020enantioselective,ali2020probing} It is clear that nonzero $\Delta\langle{F_{z,grad}}\rangle$ is achieved under certain experimental conditions. In particular, the maxima are seen to co-localize with the resonant spectral position of the electro-magnetic Mie coefficient $c_1$ of the chiral NP. The black dotted lines show the location of the peak values of $\mathrm{Im}(c_1)$. Changing $\kappa$ does not alter the spectral resonances, but instead changes the sign and magnitude of the differential force proportionally. 

A careful inspection of equation (\ref{eq:force_grad_approx}) suggests that, for any chosen $ r_{NP}$ and $\kappa$ of the tip, $\Delta\langle{F_{z,grad}}\rangle$ may approach zero when the first Kerker condition of the chiral NP is met, irrespective of helicity density of the incident light. The Kerker condition is $\alpha_{ee}=\varepsilon_0\alpha_{mm}$~\cite{kerker1983electromagnetic,lee2017reexamination,hanifeh2020helicity}, and the magnitude of the quantity $\alpha_{ee}-\varepsilon_0\alpha_{mm}$ is plotted in log scale in Figure \ref{fig:fig2}(b) as a function of radius and excitation wavelength, while $\kappa$ is fixed at $0.1$. The white dotted lines depict the conditions where the logarithm of $|\alpha_{ee}-\varepsilon_0\alpha_{mm}|$(normalized to maximum) has a value below -1.5 and the relation $\alpha_{ee}=\varepsilon_0\alpha_{mm}$ is approximately satisfied whereas the black dotted lines show the resonant position of electro-magnetic Mie coefficient $c_1$ as found in panel (a). The finite separation between the two curves assures that the Kerker condition is less likely to have any impact on  helicity density measurements if the force difference measurement is properly maximized. 


\section{Full Wave Simulations}\label{sec:FWS}
\begin{figure}[h]
\centering
\includegraphics[height=10cm]{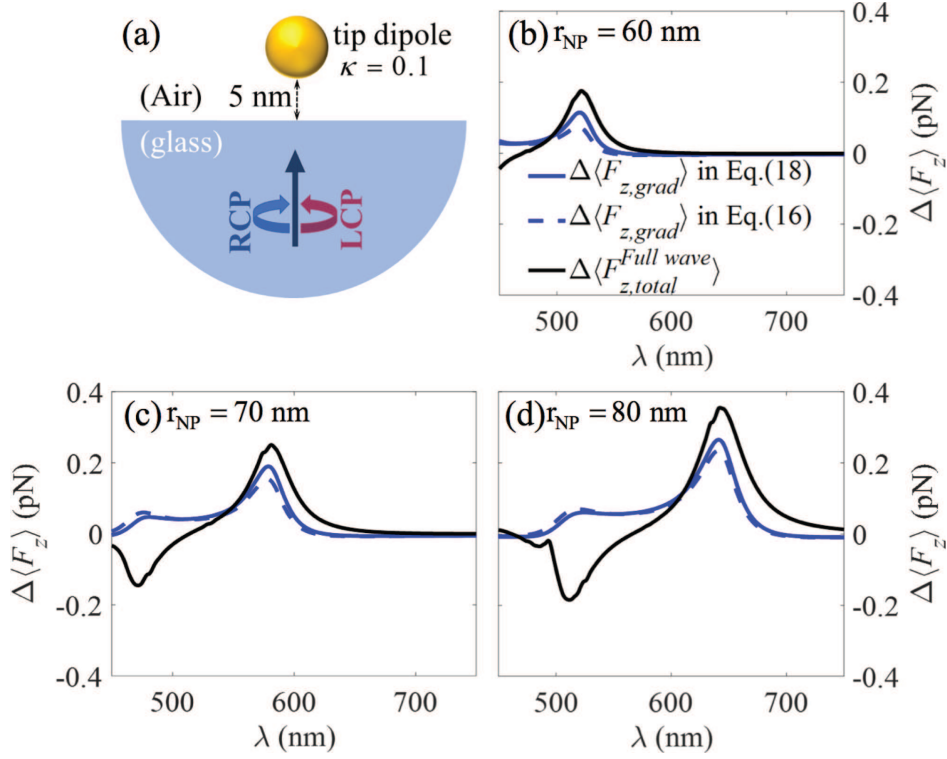}
\caption{(a) Schematic of the chiral tip sphere above a glass substrate ($n=1.5$) with radius $r_{NP}$ and tip-surface and glass surface distance of $5~{\rm{nm}}$ used in the FEM simulation. The material properties of the sphere, $\varepsilon_r$ and $\mu_r$ are the same as silicon, with $\mu_r=1$ and $\kappa=0.1$. The system is illuminated by an incoming plane wave (first LCP then RCP) from below. Time-averaged differential gradient force from analytical calculation using equation \ref{eq:force_grad_approx} (blue solid) and equation \ref{eq:force_grad} (blue dotted) and total force from full-wave simulations (black solid curve),  for a chiral NP radius with radius (b) $r_{NP}=50~\rm{nm}$, (c) $r_{NP}=60~\rm{nm}$, and (d) $r_{NP}=70~\rm{nm}$. }
\label{fig:force_FEM}
\end{figure}

To further validate our analytical findings, we perform 3D full wave simulations to determine the force exerted on a chiral tip when it is placed above a dielectric substrate. We use the finite element method implemented in COMSOL Multiphysics. In the simulation, the tip is modeled as an isotropic chiral sphere \cite{yang2016resonant, nieto2004near,PhysRevLett.99.127401} with the same material parameters as used above in Mie scattering formalism. Indeed, $\varepsilon_r$ and $\mu_r$ of the sphere are same as that of silicon with $\mu_r=1$ and the bulk chirality parameter is $\kappa=0.1$. As shown in figure \ref{fig:force_FEM}(a), the sphere is placed above a semi-infinite glass substrate  ($n=1.5$) and the  distance between the glass surface and the sphere surface is $5~\rm{nm}$. The system is then sequentially illuminated by LCP and RCP plane waves of field strength $1.5\times10^6~\rm{Vm}^{-1}$ in the bottom illumination scheme. The force exerted on the chiral sphere is determined by integrating the Maxwell’s stress tensor over the outer surface of the tip (i.e., the sphere).\cite{Yang2016}

Figures \ref{fig:force_FEM}(b), (c), and (d) show the time averaged-force difference spectrum for spheres of radius of 50~nm, 60~nm, and 70~nm, respectively. The blue solid and dotted curves show the analytical result using equation (\ref{eq:force_grad_approx}) and equation (\ref{eq:force_grad}), whereas the full-wave FEM result is indicated by the black solid line. For all three cases, the FEM simulation closely follows the analytical result obtained for the main dipolar mode, which is found at the longer wavelength and peaks with the Mie resonance of the $c_1$ coefficient. On resonance, the dipolar differential gradient force (blue curves) constitutes the dominant contribution to the total force difference (black curve). The dominance of the dipolar contribution grows more obvious as the radius is increased. The difference between the FEM simulation and the analytical results can be attributed to higher order multipole contributions, which are not considered in the dipolar approximation. The small difference between the solid and dotted blue curves is due to the contribution of $z^{-5}$ distance dependent force term in equation (\ref{eq:force_grad}). This term does not carry any information on the helicity density and its significance decreases for larger radii, as is evident in Figures \ref{fig:force_FEM}(b)-(d). Near resonance, the differential force is of the order of 0.25 pN for the 80~nm radius tip ($\kappa=0.1$). Although such forces are near the noise floor of a typical PiFM microscope, sensitive experiments are likely able to resolve the targeted differental force under optimized conditions.\cite{yamanishi2021optical}

The dip  at lower wavelengths (black curve) for the $70$~nm and $80$~nm radius tips in Figure \ref{fig:force_FEM}(c) and (d) is attributed to the differential interaction force, as defined in equation (\ref{eq:force_scat}). This is because, equation (\ref{eq:force_grad_approx}) is positive-valued whereas (\ref{eq:force_scat}) and the $z^{-5}$ dependent term of equation (\ref{eq:force_grad}) are negative-valued in the considered wavelength range. In the same lower wavelength region, the magnitude of $\Delta\langle{F_{z,grad}}\rangle$ is comparatively much smaller than the magnitude of $\Delta\langle{F_{z,int}}\rangle$ for the $70$~nm and $80$~nm radius tips. 


\section{Force Map of Helicity Density}
\begin{figure}[h]
\centering
\includegraphics[width=10cm]{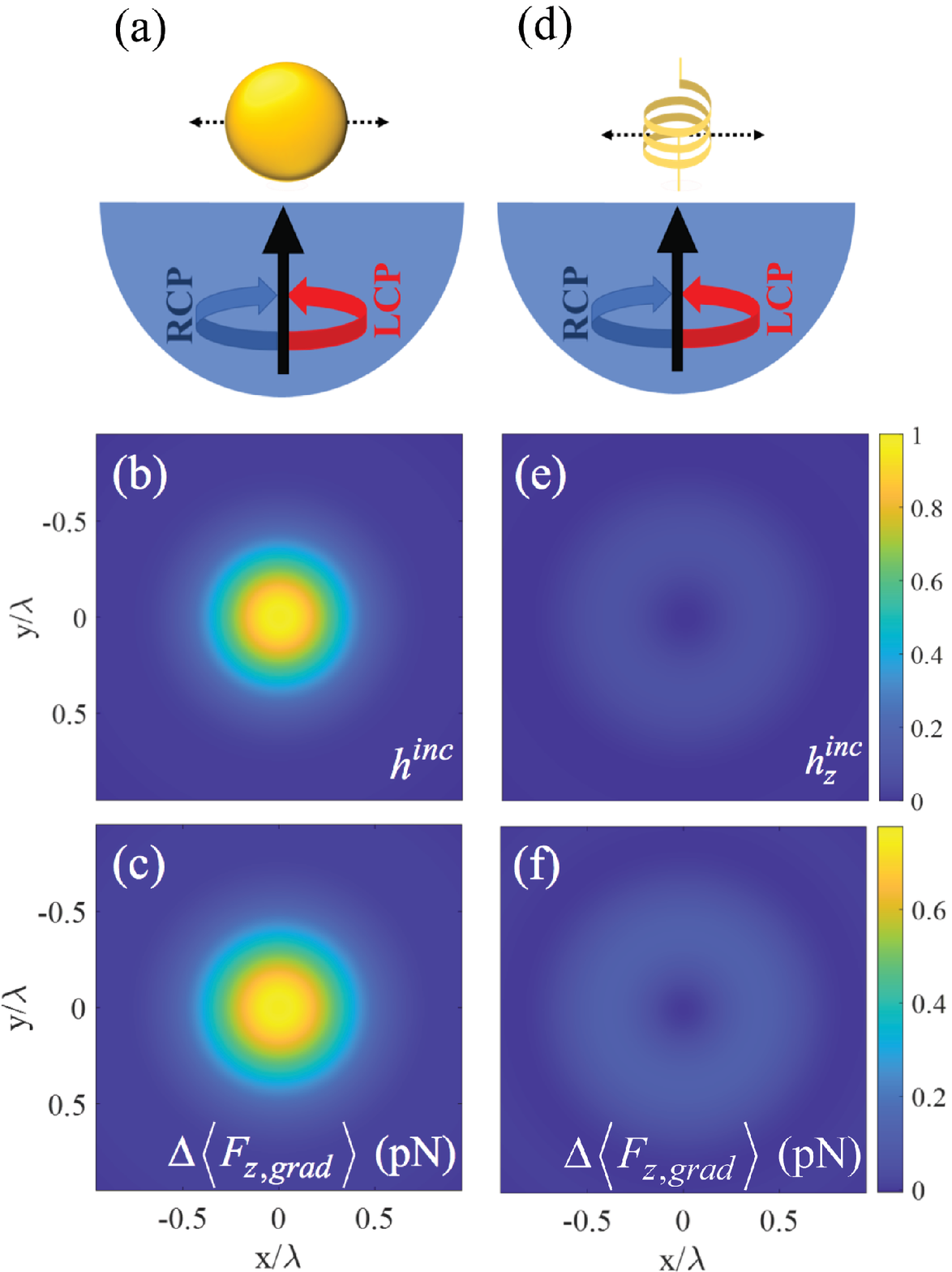}
\caption {(a) Sketch of an isotropic chiral scatterer being horizontally scanned over a glass substrate through the focus of a circularly polarized (LCP/RCP) beam. (b) The focal plane distribution of total normalized helicity density ($h^{inc}=\frac{1}{2\omega c}\rm{Im}\{\mathbf{E^{inc}\cdot\mathbf{H^{inc\ast}}}\}$) of the incident circularly polarized (LCP) light focused by a 1.4 NA oil objective. (c) The differential  force map, $\Delta\langle{F_{z,grad}}\rangle$ at the focal plane for a chiral isotropic tip of radius $r_{NP}=80~\rm{nm}$. (d) Sketch of a helical-shaped tip in the same configuration as in (a). (e) Spatial dependence of the longitudinal component of the normalized helicity density ($h^{inc}_z=\frac{1}{2\omega c}\rm{Im}\{{\mathit{E}^{inc}_z{\mathit{H}^{inc\ast}_z}}\}$) in (b). (f) Differential force map for the helical-shaped tip. The polarizability strength of the isotropic and helical-shaped tip has been set to the same value.}
\label{fig:force_map}
\end{figure}

Finally, we present the differential force  map of the helicity density for a focused chiral beam, a scenario of direct relevance to microscopy applications. The situation is schematically shown in Figure \ref{fig:force_map}(a) where an isotropic chiral sphere of radius $r_{NP}=80~\rm{nm}$ is scanned over the focal region of the incident beams above glass substrate. We assume a circularly polarized Gaussian beam of $2~\rm{mW}$ average power at a wavelength of $\lambda=641~\rm{nm}$ as the incident light source, which is focused by a 1.4 NA oil~($n=1.518$) objective lens. The focal plane electric and magnetic field components used in the simulation are obtained from reference [32], page 62. We first calculate the normalized helicity density distribution at the focal plane for a focused LCP beam, which is shown in \ref{fig:force_map}(b). We next calculate the differential force $\Delta\langle{F_{z,grad}}\rangle$ using equation (\ref{eq:force_grad_approx}). In Figure \ref{fig:force_map}(c), $\Delta\langle{F_{z,grad}}\rangle$ is shown for the case of a tip of radius $r_{NP}=80~\rm{nm}$. As equation (\ref{eq:force_grad_approx}) suggests, the force difference has the same spatial dependence as the non-zero helicity density of the focused LCP beam shown in Figure \ref{fig:force_map}(b). 

Last, we consider a realistic chiral nanoprobe obtained through helical carving of regular achiral tips.\cite{zhao2017nanoscopic} Such chiral tips display dominant longitudinal chirality\cite{caloz2020electromagnetic} and in the dipolar approximation this is reflected in the polarizabilty tensors as\cite{caloz2020electromagnetic,A}\\
\begin{equation}
\begin{split}
~&{\underline{\boldsymbol{\alpha}}}_{ee}= \alpha_{ee}^{zz}\hat{\mathbf{z}}\hat{\mathbf{z}};\\
&{\underline{\boldsymbol{\alpha}}}_{mm}= \alpha_{mm}^{zz}\hat{\mathbf{z}}\hat{\mathbf{z}}\\
&{\underline{\boldsymbol{\alpha}}}_{em}=\alpha_{em}^{zz}\hat{\mathbf{z}}\hat{\mathbf{z}}
\end{split}
\label{eq:polarizability_helix}
\end{equation}
Here we have ignored the non-diagonal components of the polarizability tensors for brevity. Under these conditions, the differential gradient force for the helical-shaped tip turns out be 
\begin{equation}
\begin{split}
\Delta\langle{F_{z,grad}}\rangle\sim~&\frac{6\omega c}{\pi|\mathit{z}|^4}
\mathrm{Im}\Big\{\frac{\alpha^{zz\ast}_{em}}{\varepsilon_0}(\alpha_{ee}^{zz}-\alpha^{zz}_{mm}{\varepsilon_0})\Big\}  \mathit{h}^{inc}_{z}\end{split} \label{eq:force_grad_approx_helix}
\end{equation}
The details of the derivation can be found in Supporting Information (SI Section 3). Interestingly, the differential gradient force for the helical-shaped tip is only sensitive to the longitudinal component of the helicity density. In contrast, the isotropic chiral tip measures both the transverse and longitudinal components in the differential force measurement, as expressed in equation (\ref{eq:force_grad_approx}). 

To demonstrate these features, we consider a helical-shaped tip that is placed above a glass substrate, and  subsequently scanned over the focal region as sketched in Figure \ref{fig:force_map}(d). Figure \ref{fig:force_map}(e) shows the calculated longitudinal component of the helicity density, revealing a donut shaped profile. Next, to calculate the differential force we use the same polarizability values of the isotropic chiral sphere in Figure \ref{fig:force_map}(a) for $\alpha_{ee}^{zz}$, $\alpha_{mm}^{zz}$ and $\alpha_{em}^{zz}$. The map of $\Delta\langle{F_{z,grad}}\rangle$ for the helical-shaped tip is depicted in figure \ref{fig:force_map}(f), showing the expected donut profile that replicates the longitudinal component of the helicity density of the incident LCP beam. In the Supporting Information, we also present the helicity density maps determined through the differential force of a focused azimuthally radially polarized beam (ARPB)\cite{kamandi2018unscrambling,hanifeh2020optimally,novotny2012principles}.


\section{Conclusion} We have studied the information contained in the photo-induced force exerted on a chiral tip (modeled as a chiral nanosphere) when it is illuminated by chiral light. Our theoretical analysis reveals that the differential force is directly sensitive to the chiral properties of light. In particular, the dominant component to the differential gradient force is directly proportional to the helicity density of the incident chiral light, whereas the differential scattering force is sensitive solely to the spin angular momentum of the applied light. Using realistic values for the illumination intensity, tip dimension, and the chirality parameter of the tip, we find that the differential force can reach detectable values of several hundreds of fN, just above the noise floor of common scan probe microscopy systems. These findings are significant because a direct characterization of optical chirality at the nanoscale has hitherto been extremely challenging. The observation that the differential gradient force can map out the local helicity density of the light is relevant for numerous applications where knowledge of the chiral state of light at sub-diffraction-limited dimensions is important.

\section*{acknowledgement}
\noindent The authors acknowledge support by the W. M. Keck Foundation, USA, and the National Science Foundation, grant CMMI-1905582.

\section*{Supporting Information}
\noindent Detailed derivations are available in the Supporting Information. Please contact the corresponding author for a copy.

\bibliography{library}

\end{document}